\documentclass[final,3p,times]{elsarticle}
\makeatletter
\def\ps@pprintTitle{%
  \let\@oddhead\@empty
  \let\@evenhead\@empty
  \def\@oddfoot{\reset@font\hfil\thepage\hfil}
  \let\@evenfoot\@oddfoot
}
\makeatother
\usepackage{amsmath}
\usepackage{amssymb}

\usepackage{amsfonts}
\usepackage{mathrsfs}

\usepackage{graphicx}
\usepackage{dcolumn}
\usepackage{bm}
\usepackage{epsfig}
\usepackage{enumerate}
 \biboptions{sort&compress}
 \usepackage{float}
\usepackage{amssymb}

\def\be{\begin{equation}}
\def\ee{\end{equation}}
\def\bea{\begin{eqnarray}}
\def\eea{\end{eqnarray}}
\def\bfl{\begin{flushleft}}
\def\efl{\end{flushleft}}
\def\bfr{\begin{flushright}}
\def\efr{\end{flushright}}
\def\bc{\begin{center}}
\def\ec{\end{center}}
\def\ben{\begin{enumerate}}
\def\een{\end{enumerate}}
\def\bit{\begin{itemize}}
\def\eit{\end{itemize}}

\def\dzn{,\kern-0.1em,}

\def\d#1{{#1\kern-0.4em\char"16\kern-0.1em}}
\def\D#1{{\raise0.2ex\hbox{-}\kern-0.4em 31}}

\newcommand{\tm}[1]{\mathrm{#1}}
\newcommand {\apgt} {\ {\raise-.5ex\hbox{$\buildrel>\over\sim$}}\ }
\newcommand {\aplt} {\ {\raise-.5ex\hbox{$\buildrel<\over\sim$}}\ }

\journal{Computer Physics Communications}

\begin{document}

\begin{frontmatter}



\title{Multipath Metropolis Simulation of Classical Heisenberg Model}


\author[ftn]{Predrag S. Raki\' c\corref{cor1}}
\ead{pec@uns.ac.rs}

\author[pmf]{Slobodan M. Rado\v sevi\' c}
\ead{slobodan@df.uns.ac.rs}

\author[pmf]{Petar M. Mali}
\ead{petar.mali@df.uns.ac.rs}

\author[ftn]{Lazar M. Stri\v cevi\' c}
\ead{lucky@uns.ac.rs}

\author[ftn]{Tara D. Petri\' c}
\ead{tara.petric@gmail.com}

\cortext[cor1]{Corresponding author}

\address[ftn]{Faculty of Technical Sciences, University of Novi Sad,
Trg Dositeja Obradovi\' ca 6, 21000 Novi Sad, Serbia}
\address[pmf]{Department of Physics, Faculty of Science, University of Novi Sad,
Trg Dositeja Obradovi\' ca 4, 21000 Novi Sad, Serbia}

\begin{abstract}
Processor cores are becoming less expensive and thus more accessible.
To utilize increasing number of available computing elements, good 
parallel algorithms are necessary.
In light of these changes in contemporary computing,
multipath Metropolis simulation of classical Heisenberg model is explored.
In contrast to the original single--path algorithm, 
multipath simulation approach is inherently parallel
because different random--walk paths are mutually independent.
This independence enables easy and efficient harnessing of numerous cores' computing power
in embarrassingly parallel algorithms.
Aside form being inherently parallel, multipath simulation approach results
in independent and normally distributed simulation output.
Normal distribution enables simple and straightforward statistical processing.
Thus, multipath simulation results can be easily computed with
arbitrary and statistically known precision.

\end{abstract}


\end{frontmatter}

\section {Introduction}
Contemporary evolution of Monte Carlo (MC) methods in statistical physics
is dictated by several important issues.
These include (but are not limited to) the development of faster and more efficient algorithms, 
further improvement of  (pseudo)random number generators, reduction
of correlation between data, nontrivial statistical analysis as well as 
construction of more powerful computing machines
(see e.g \cite{Binder12,Cep,Binder} and references therein). 
A progress in each of uppermentioned fields is welcomed
since variety of possible applications may put emphasis on different aspects of
simulation  and can also initiate the advance in other areas.
In the present paper we offer a strategy  for circumventing the complications with
statistical analysis of correlated data in the case of classical
Heisenberg model (CHM). 
Even though standard methods for dealing with correlated data,
such as blocking or jackknife method \cite{Henrik,Ferrenberg91,PRD,Morales}, do exist, 
it is not hard to imagine the situation in which truly independent
simulation results are desirable. 
For example,  the jackknife method is known to  fail in some cases
\cite{Jack}.
Being used in many different areas 
of physics as a working model (see  
\cite{Justin,EuX,EuX2,Stenli,ptice,PhysicaA,CPC2} and references therein),
CHM is perfectley suited for present exposition.
Analogus situation holds for the Ising model (IM), which is still
atractive for MC simulations, as seen from recent papers \cite{GPU1,GPU2,GPU3,FPGA}.
We shall brieflly discuss the simulation of  IM on the square lattice from the present point of view 
in   \ref{ising}.
Current work is stimulated by recent developments in the hardware industry:
over the last couple of decades the price of processing elements has constantly decreased.
This trend is, most likely, set to continue for at least another decade more.
Hardware manufacturers have turned to multicore systems \cite{Geer2005},
resulting in computer architectures with more processing elements. 
Although significantly faster cores are not expected at least in the near future, 
a rapid increase in the number of computing cores is a reasonable expectation.

When single core computers were dominant,
the main concern was efficient utilization of memory and processor cycles.
These aspects are still important however,
the increasing number of available cores has also led
to a new goal: to utilize as many computing cores as possible
in efficient and scalable parallel algorithms.
Often there is a trade--off, since increasing the number of utilized cores cannot be achieved without
a significant increase in the total number of computing cycles.
Thus, it is necessary to develop algorithms that utilize available 
processing elements efficiently. 
In this context an inherently parallel multipath approach
corresponds to contemporary hardware and software trends.

A Metropolis algorithm is originally defined  as 
a single--flip (only one particle moves at a time),
single--path (a complete simulation is performed on one, long
random--walk path) algorithm \cite[p.~1088]{Metropolis53}.
It is an inherently sequential modification of the Monte Carlo scheme,
because each configuration depends on the previous one.
One value (of some system property) is computed out of the system state after each flip.
At the end, all values are averaged.
The mean of these values is taken as the simulation result.
It is quite an efficient and widely applicable algorithm,
though it produces values that are 
correlated \cite{Ferrenberg91} and,
because of that, only asymptotically normally distributed \cite{Binder2,FPGA}.
Such distribution complicates statistical analysis. 
Also, convergence speed is issue in a single--path approach,
since it can be prone to entrapment 
in local minima \cite{Altekar}\footnote{The paper actually refers to more 
general category of algorithms called Markov chain Monte Carlo (MCMC).}.
To overcome this issue, algorithm improvement that includes more than one 
random walk path, known as a Metropolis coupled Markov chain Monte Carlo $(MC)^3$
is proposed in \cite{geyer1991MC3} (see also \cite{Geyer2}). 
A decades old idea of more--than-one--path Metropolis algorithm enhancement
is extended in this paper into an embarrassingly parallel and highly scalable 
multipath simulation approach, 
appropriate for our highly parallel computing age.

Section \ref{SectSim} gives an overview of the multipath (MP)
Metropolis simulation for the CHM. The notions of simulation path (SP) and
simulation output (SO), used throughout the text, are defined there.
Section \ref{SectSimOut} contains the discussion on the quality
of the output data in the MP simulation. As the output data are uncorrelated by
the very construction of the MP simulation, the subsequent
statistical analysis is not only greatly simplified, but also made rigorous.
Thus, the results presented in  Section \ref{SectSimOut} reveal
the most obvious advantage of the MP simulation over standard (i.e. single-path) one.
The conection betveen MP and single-path simulation of the CHM is
provided in the section \ref{SectResult}, with emphasis on the order
parameter (spontaneous magnetization). By computing various thermodynamic
properties of CHM in MP and single-path simulation, we demonstrate an
overall agreement.
Finally, the IM on a square lattice is discussed in  \ref{ising}, where we 
test MP simulation against standard single-path approach and the
Onsager-Yang solution.
Results obtained  for  models vith continous  (CHM)
and descrette internal symmetry (IM) give us the reason to bellive that 
increment in precission/accuracy is possible and that
MP simulation could be  of use in other cases
where standard Markov-chain MC methods apply.

The simulations on large lattices ($L>10$) for CHM were conducted using data storage and computing services of 
the Supercomputing  Center of Galicia (CESGA) \cite{Cesga}. 
We used
the FINISTERRAE \cite{CesgaFINISTERRAE} (2500 cores, 35 TFLOPS) and SVG \cite{CesgaSVG} 
(1800 cores) supercomputers running under GNU/Linux operating system. All simulation results presented in this paper were 
computed using free software C++ library called ''Hypermo'' \cite{web-Hypermo}, while
 most of the figures are created with ''Tulipko'' \cite{web-tulip}
interactive visualization tool. More details on "Tulipko" will be
presented elsewhere.

\section {Simulation} \label{SectSim}

In this section we define the model, quantities of interest
and set up notation to be used throughout the paper. 
We also specify the  way in which simulation is conducted. 
In order to acquire statistically independent values, 
to overcome the issue of local minima entrapment
and to be able to utilize the now abundant computing cores,
the multipath (MP) approach is explored.
It produces a normally distributed simulation output, 
in both  ferromagnetic and paramagnetic phase
with simulation output values statistically independent by definition.
The statistical analysis of data set with such distribution characteristics 
is straightforward and rigorous, i.e. there is no need for  approximations 
or additional calculations
inevitable when working with asymptotic distributions and correlated data.
In this approach a Metropolis algorithm 
is applied on many different paths,
all beginning at a randomly chosen state and 
each path producing only one simulation output (after reaching thermal equilibrium).
Beside producing independent output values
this approach offers much bigger parallelization potential compared to 
single--path approach, since computations on each path are mutually independent.

\subsection{Classical $O(3)$ Heisenberg model}

Classical $O(3)$ Heisenberg model is defined by the Hamiltonian (energy function) 
\begin{equation}
H=-\frac{J}{2}\sum_{\bm{n},\bm{\lambda}}\bm{S}_{\bm{n}}\cdot\bm{S}_{\bm{n}+\bm{\lambda}}
\label{ham},
\end{equation}
where $J$ is constant of coupling between nearest neighbors (nn)
spins and summation is taken over all lattice sites $\{\bm{n}\}$, and $\bm{\lambda}$ 
connects a given site to its nearest neighbors.
 For $J>0$ the ground state
is ferromagnetic, and for $J<0$ antiferromagnetic, but this
makes no difference at the classical level.
We define the
energy scale by fixing the exchange integral $J=1$. The set
$\{\bm S_{\bm n}\}$ represent spin vectors on a 
periodic simple
cubic lattice with $N=L^3$ sites, whose
position is defined by the totality of $\{\bm n\}$.
We employ the standard spherical parametrization
 \be \bm{S}_{\bm{n}}=[\sin \theta_{\bm{n}} \cos \varphi_{\bm{n}},\sin 
\theta_{\bm n}\sin \varphi_{\bm{n}},\cos \theta_{\bm{n}}]^{\mathsf{T}}
\ee
for unit vectors.
\subsection{Definitions and notation}

Metropolis single spin--flip algorithm is used to simulate a 3-dimensional
classical Heisenberg on a simple cubic lattice. 
All simulations discussed in this paper are conducted 
on a particular lattice size at a particular temperature.
Each simulation consists of a certain number of simulation paths 
(simulation path, SP).
Each SP produces output which consists of four values
$\{M_i^x,M_i^y,M_i^z,H_i\}$, where index
 $i=1,2,\dots, \mathcal{N}$ labels the simulation paths. The first three
components in each output define the instantenous
total spin per lattice site, $\bm M =  (1/L^3)\sum_{\bm n} \bm{S}_{\bm n}$, while the fourth one
is the instanteous energy per lattice site.

Outputs of all the $\mathcal{N}$ SPs, together, form a simulation output (SO).
SO that meets desired precision and accuracy is called representative.
Additional statistical and physical computations on representative SO
are necessary in order to obtain particular physical values,
these are referred to as simulation results (simulation result, SR).

Each SP starts from a randomly chosen state which corresponds to 
infinitely high temperature state \cite{Binder2010}. 
Initial state is selected by randomizing the states of all lattice sites 
i.e. by randomly choosing values of both angles at all lattice sites. 
After the initial randomization,
Metropolis algorithm is applied to a number of sites
in an attempt to reach some thermal equilibrium state of the lattice. 
Execution of Metropolis algorithm on $L^{3}$ randomly chosen lattice
sites is referred to as one lattice sweep (LS) \cite{Binder}.
In every SP a number of LSs is conducted.
The SP output is computed out of the last lattice state 
i.e.\ from the final state of the Markov chain
(state after the last LS). The MC averages are  then computed as
\be
\langle A \rangle = \frac{1}{\mathcal N} \sum_{i = 1}^{\mathcal N} A_i \label{MCAV}
\ee
Specifically, the magnetization and the internal energy are defined as SO average values
\be \langle \bm{M} \rangle=
\frac{1}{{\mathcal N}}\sum_{i=1}^{\mathcal N}\bm{M}_i, \hspace{1cm} \langle H \rangle =  
\frac{1}{\mathcal N}\sum_{i=1}^{\mathcal N} H_i
\label{M_and_H}.
\ee
We also make use of 
\be \langle M^{k} \rangle=\langle |\bm{M}|^k \rangle= 
\frac{1}{{\mathcal N}}\sum_{i=1}^{\mathcal N}({M}_i)^k , 
\hspace{1cm} M_i=\sqrt{(M^x_i)^2+(M^y_i)^2+(M^x_i)^2}
\ee
for $k=1,2,4$. In particular, $\langle M \rangle$ plays the role of order parameter
on finite lattices. We shall further consider susceptibility and heat capacity
\be \chi(T)=\frac{L^3}{T} \left[\langle |\bm M|^2 \rangle-\langle |\bm M| \rangle^2 \right],
\hspace{1cm}
C_V(T)=\frac{L^3}{T^2}\left[\langle H^2 \rangle-\langle H \rangle^2 \right].\label{cv}
\ee 
All thermodynamic quantities, obtained   both in standard single-path and multipath method,
will be calculated per lattice site.
Finally, the transition temperature for infinite lattice is obtained from the
intersection of fourth order cumulants \cite{Binder}
\begin{equation} 
\label{kum} U_L=1-\frac{1}{3}\frac{\langle M^4 \rangle}{\langle M^2 \rangle^2} 
\end{equation}
for different linear finite lattice dimensions $L$.

\section{Simulation output} \label{SectSimOut}

As already stated, the MP simulation is designed so that it
produces statistically independent output.
To confirm SO independence, output components distribution is carefully examined
in  ferro and para phase. 

\begin{figure}[ht] 
\includegraphics{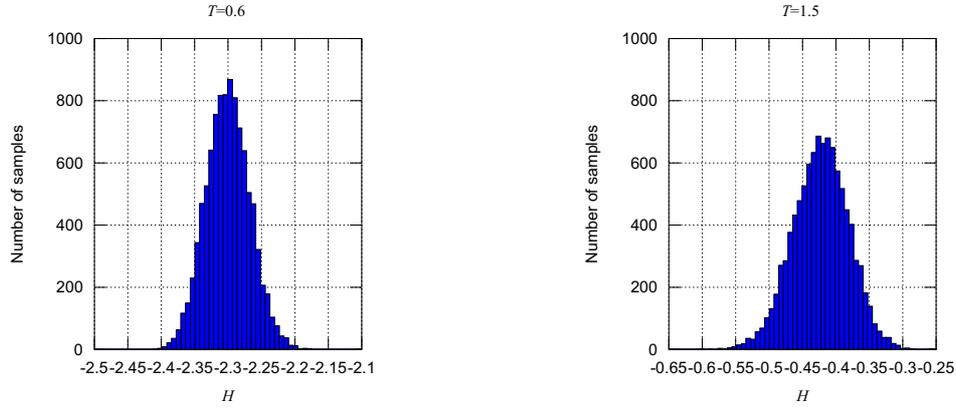}
\centering
\caption{(Color online) Energy distribution histogram --- ordered and disordered phase ($L=10$)}
\label{Fig:Energy_distribution}
\end{figure}

%
Figure~\ref{Fig:Energy_distribution} 
shows SO energy distribution 
in both the ordered phase ($T=0.6$) and the disordered phase ($T=1.5$).
In both phases energy is normally distributed, confirming SO independence.
SO energy values $\langle H \rangle$ accumulate around internal energy per site (median).
We observed that energy is a little more dispersed in the para phase
(the right histogram is a little lower and wider).

\begin{figure}[pht] 
\includegraphics{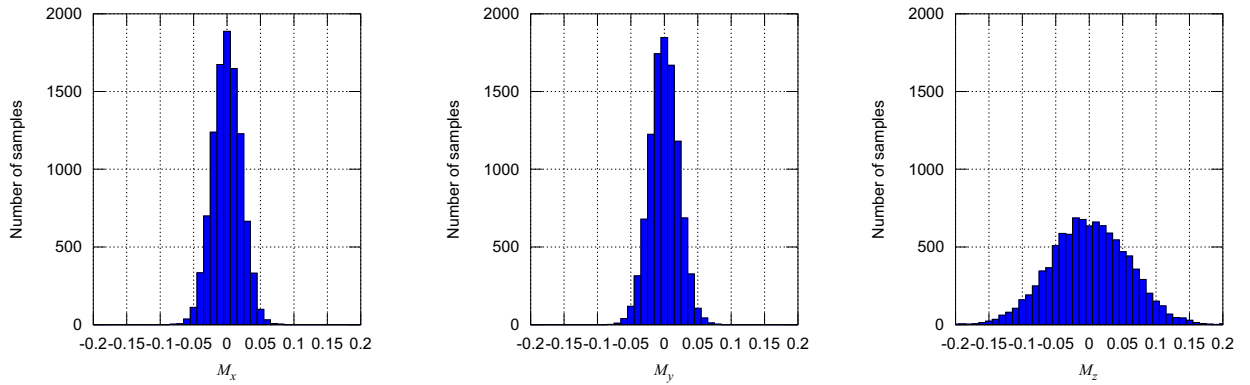}
\caption{(Color online) Total spin components distribution --- histogram --- 
disordered phase ($L=10$, $T=1.5$)}
\label{Fig:Magnetisation_hist_high}
\end{figure}

Likewise, Figure~\ref{Fig:Magnetisation_hist_high}
shows that the distributions of all three total spin components  in the 
disordered phase are normally distributed, accumulating around zero.
The $M_z$ component is significantly more dispersed than the other two components
since temperature $T=1.5$ is not far from the ordered phase. 
$M_z$ dispersion decreases as the lattice goes deeper into the disordered phase.

Normal Q--Q plot (sometimes also called normal probability plot) 
is a graphical method for assessing whether or not 
a data set is approximately normally distributed.
It is a special case of Q--Q plot, in which normal distribution is plotted on the $x$--axis.
In normal Q--Q plot a data set is plotted
in such a way that the points should form the straight line 
if normal distribution is a good fit for the set.
\begin{figure}[pht] 
\includegraphics{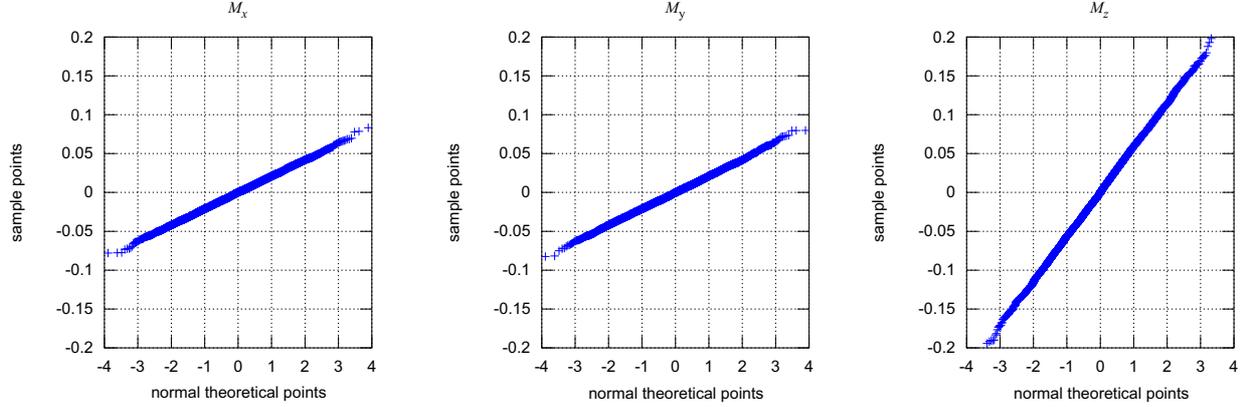}
\caption{(Color online) 
Total spin components distribution --- Q--Q plot --- disordered phase. 
Inverse standard normal distribution, i.e. normal theoretical points, are plotted on $x$-axis.
($L=10$, $T=1.5$)
}
\label{Fig:Magnetisation_qqplot_high}
\end{figure}
Figure~\ref{Fig:Magnetisation_qqplot_high}
shows the distributions of all three total spin components,
similar to Figure~\ref{Fig:Magnetisation_hist_high},
but this time as normal Q--Q plots.
It is obvious that points can be neatly fitted to the line.

For quantitatively measuring the goodness of the fit,
the second norm of the fit residuals can be used \cite[p.~210]{hansen2011octave}:
\be
norm(R,2)=\sqrt{\sum_{i=1}^N R_i^2};
\qquad
R_i = X_i - f(i)
\ee
where $R$ represents fit residuals set,
$X$ is fitted data set and $f$ is fitting function.
Unfortunately, norm depends on the absolute values of the residuals
and as such cannot be used for comparison between different fits.
Instead, 
for comparison between samples with different means, correlation coefficient 
can be used.
It will be referred to as distribution coefficient and is defined as:
\be
r=1-\frac{norm(R,2)^2}{(N-1)\sigma^2}; \qquad 0\leq r\leq1
\label{dist-coefficient}
\ee
where $R$ represents fit residuals set,
$N$ is number of fitted points (number of elements in the $X$ set) and 
$\sigma$ is standard deviation of the fitted data set $X$.
Closer the value of $r$ is to one, closer the sample distribution is 
to the perfect normal distribution.
Distribution coefficients for magnetization components $M_x$, $M_y$ and $M_z$ 
denoted as $r_{M_x}$, $r_{M_y}$ and $r_{M_z}$ respectively for number 
of temperatures are shown in Table~\ref{goodness-of-the-fit}. 
The table shows that distribution coefficients in ordered phase 
are very close to 1, confirming that total spin components' distributions 
are very close to the normal one.
\begin{table}
\begin{center}
\begin{tabular}{llllll}
$T$ & $r_{H}$ & $r_{M_x}$ & $r_{M_y}$ & $r_{M_z}$ & $r_{|M_z|}$\\ \hline 
0.6 & 0.9994 & 0.9998 & 0.9998 & ---    & 0.9990\\
0.8 & 0.9988 & 0.9990 & 0.9987 & ---    & 0.9983\\
1.0 & 0.9994 & 0.9999 & 0.9998 & ---    & 0.9951\\
1.2 & 0.9979 & 0.9983 & 0.9989 & 0.9856 & ---\\
1.4 & 0.9988 & 0.9998 & 0.9998 & 0.9998 & ---\\
1.6 & 0.9995 & 0.9999 & 0.9999 & 0.9999 & ---\\
\end{tabular}
\end{center}
\caption{
$L=10$. 
Distribution coefficient $r$ is number from $0$ to $1$
which describes how close are points plotted on Q--Q plot to the straight line 
i.e.\ how close is data sample distribution to the normal distribution.}
\label{goodness-of-the-fit}
\end{table}

\begin{figure}[htb] 
\includegraphics{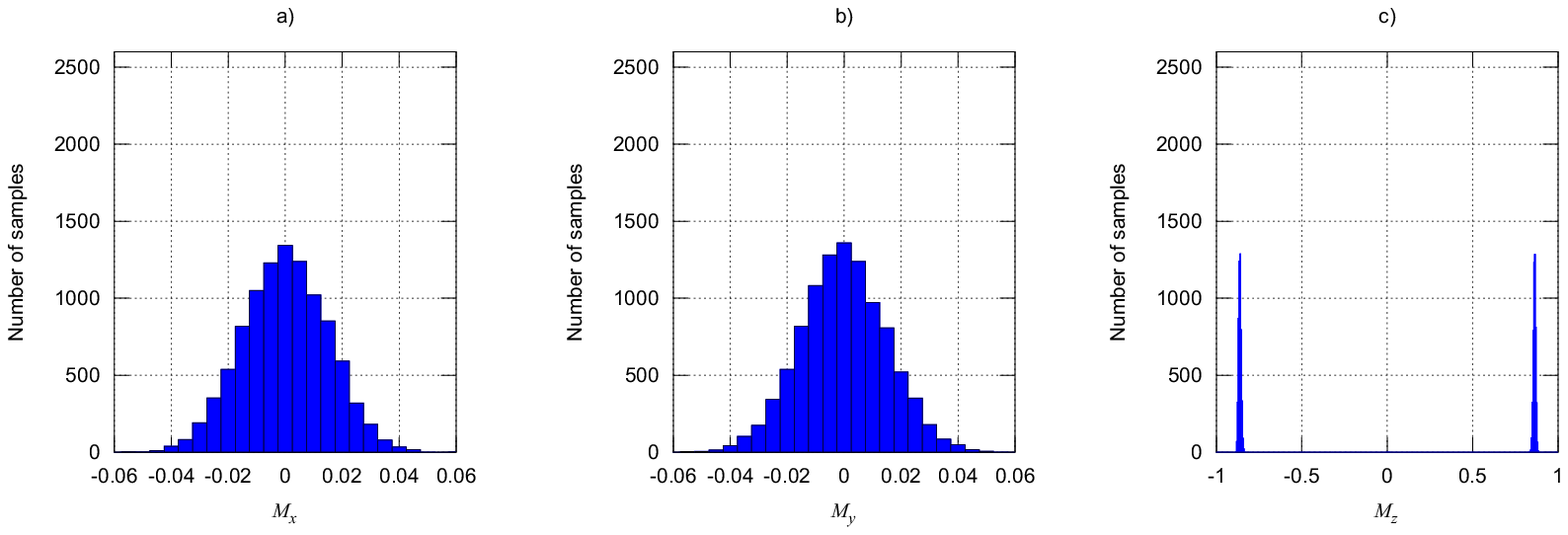}
\vskip 0.5cm
\includegraphics{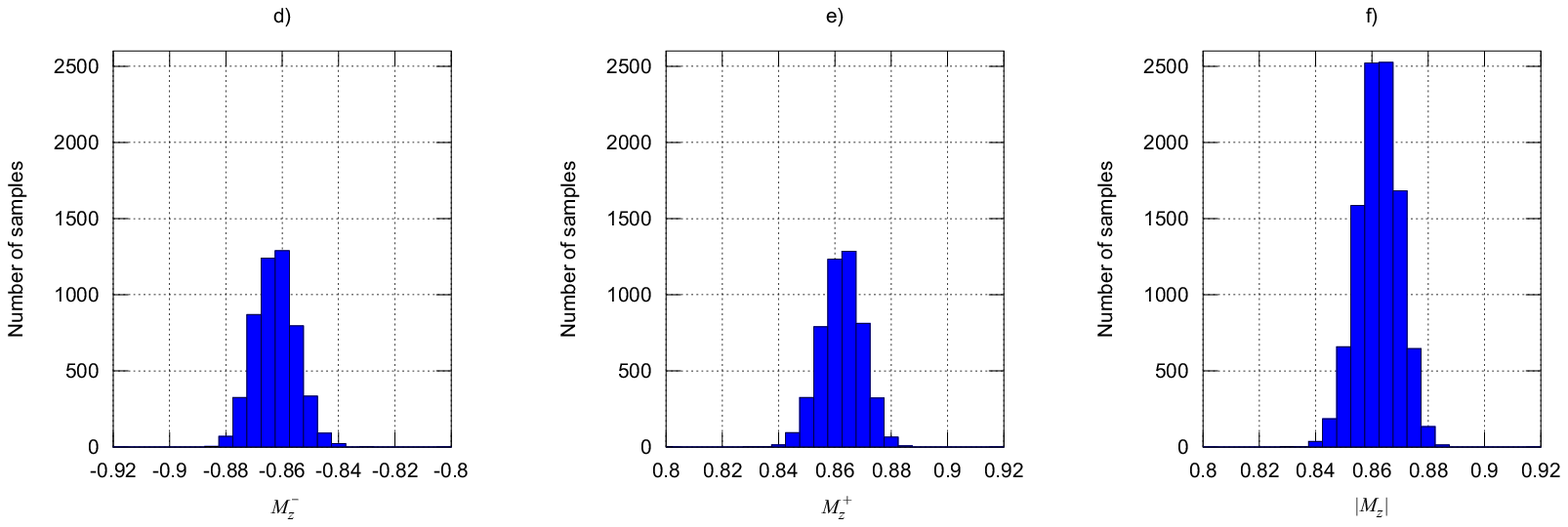}
\caption{(Color online) Total spin components distribution --- histogram --- 
ordered phase ($L=10$, $T=0.6$). d) and e) shows enlarged picks from c).}  
\label{Fig:Magnetization_hist_lo}
\end{figure}

Total spin components distributions in the ferromagnetic phase
are shown in  Figure~\ref{Fig:Magnetization_hist_lo}. It is seen that
$M_x$ and $M_y$ are normally
distributed, as in the paramagnetic phase. Note that $M_x$ and
$M_y$ group around $\langle M_x \rangle=\langle M_y \rangle=0$, in
both the ferro and para phases.
$M_z$ values form two independent normal distributions 
(see, e.g., \cite{Binder81} for the case of the Ising model),
 which reflects the
symmetry $S_{\bm{n}}^z \rightarrow -S_{\bm{n}}^z $ 
of the Hamiltonian (\ref{ham}).
As a consequence of this symmetry,  magnetization
 (defined in (\ref{M_and_H})) is
null vector. This illustrates the well known fact that there is no 
spontaneous symmetry breaking on finite lattices.

\begin{figure}[htb] 
\includegraphics{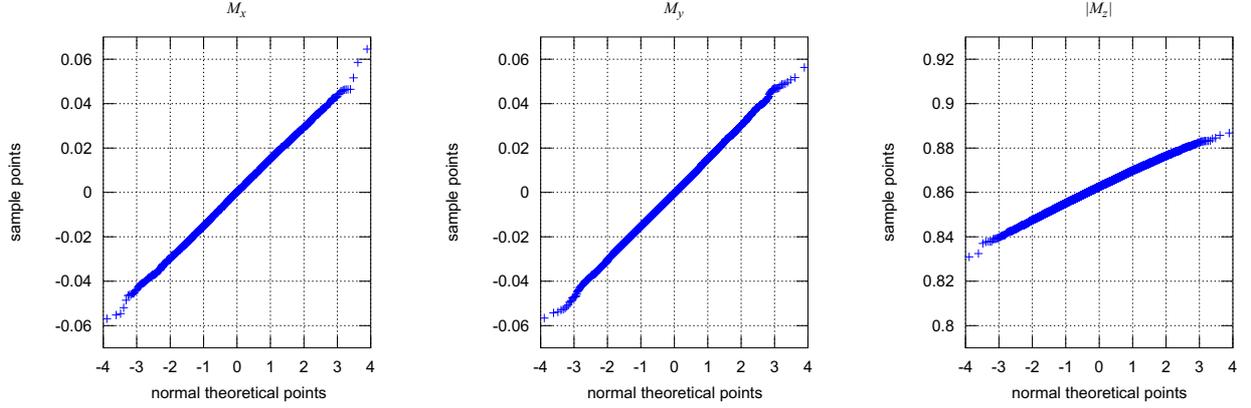}
\caption{(Color online) 
Total spin components distribution --- Q--Q plot --- ordered phase. 
Inverse standard normal distribution, i.e. normal theoretical points, are plotted on $x$-axis.
($L=10$, $T=0.6$)
}
\label{Fig:Magnetization_qqplot_lo}
\end{figure}
Figure~\ref{Fig:Magnetization_qqplot_lo} shows distributions of 
$M_x$, $M_y$ and $|M_z|$ total spin components in an ordered phase on Q--Q plot.
As on histograms (Figure~\ref{Fig:Magnetization_hist_lo}), normal distribution 
of components is observed but $|M_z|$ component dispersion is smaller 
compared to dispersions of the other two components.
In the disordered phase there is just one accumulation point for $M_z$ values 
similarly to $M_x$ and $M_y$. 
Thus, distribution coefficient for the $M_z$ component is $r_{M_z}$ 
(Table~\ref{goodness-of-the-fit}).
In the ordered phase $M_z$ values accumulate symmetrically around two points: 
$M_z^-$ and $M_z^+$, forming two normal distributions. 
In order to determine the distribution coefficient, these two sets of points 
are combined in one absolute values set.
Thus, the distribution coefficient for $M_z$ in this phase is $r_{|M_z|}$.
Trying to correlate the distribution of signed $M_z$ values to the single normal one 
in the ordered phase is meaningless.
Similarly, there is no correlation between the absolute $M_z$ values and 
normal distribution in the disordered phase.
Thus, the $r_{M_z}$ values in the ordered phase are marked with '---'
as are the $r_{|M_z|}$ values in the disordered phase.
Values from the table confirm an almost perfect fit to the straight line and 
hence to the normal distribution for energy as well as 
for all three total spin components in both the ordered and the disordered phase.

\begin{figure}[ht]
\centering
\includegraphics[width=4in]{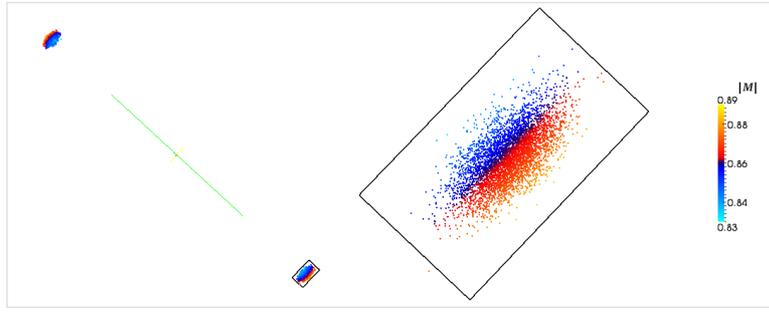}
\caption{(Color online) Total spin distribution in 3D ($L=10, T=0.6$). 
Red--blue color boundary indicates vectors with $\langle{M}\rangle$ intensity. 
Green line indicates the direction of $z$-axes. 
The midpoint of green line denotes the origin in the internal space 
of total spin vectors. 
}
\label{Fig:Magnetization_3D_T60}
\end{figure}

\begin{figure}[ht]
\centering
\includegraphics[width=4in]{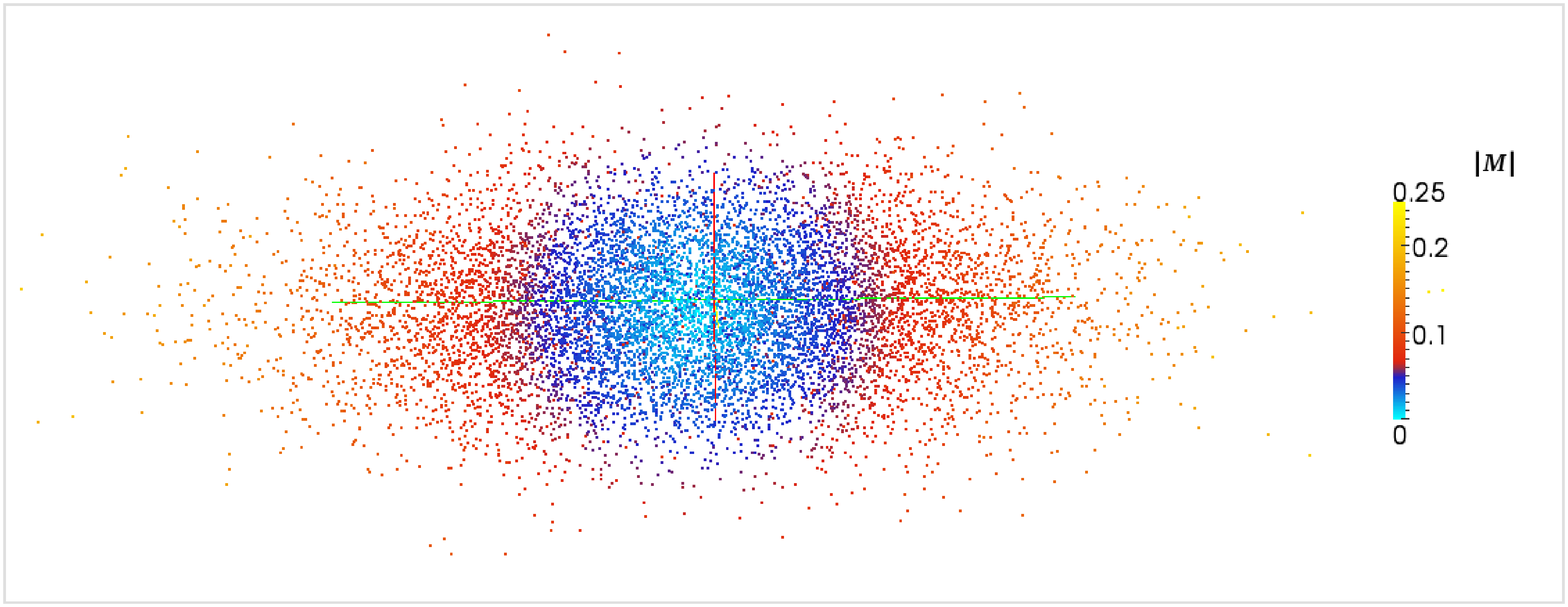}
\caption{(Color online) Total spin distribution in 3D ($L=10$, $T=1.5$).  
Red--blue color boundary emphasizes sphere with $\langle{M}\rangle$ radius. 
Green line indicates the direction of $z$-axes. 
The midpoint of green line denotes the origin in the internal space 
of total spin vectors. 
}
\label{Fig:Magnetization_3D_T150}
\end{figure}
\begin{figure}[htb] 
\includegraphics[scale=0.20]{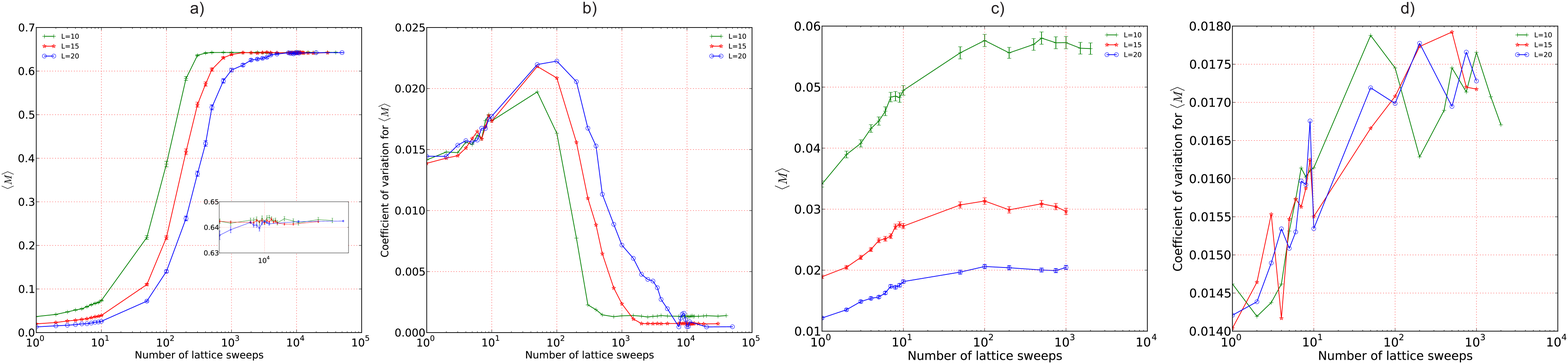}
\centering
\caption{(Color online) 
a) The order parameter $\langle{M}\rangle$ and b) coefficient of variation 
(\ref{eq:cv}) (ordered phase, $L=10, T=1.0$) 
together with c) the order parameter
and d) the coefficient of variation (disordered phase, $L=10, T=1.5$) against the
number of LSs  for $L=10, 15, 20$. $\mathcal{N}= 10^3$ both in the ordered and 
disordered phase. When not shown, the error bars are smaller than  symbols.
The inset on first graph shows a close-up of the convergence region.}
\label{ConvLatticesABCD}
\end{figure}

Figure~\ref{Fig:Magnetization_3D_T60} shows the total spin vectors 
intensity and orientation in space at a low temperature ($T=0.6,L=10$). 
The initial point of all vectors is located at the origin and
the colored dots represent terminal points of the vectors.
Vectors  group around the $z$-axis in two bundles, as
already discussed (Figure~\ref{Fig:Magnetization_hist_lo}).
In contrast, in the disordered phase spins are distributed in
the subspace of internal space around the origin,
as shown on Figure~\ref{Fig:Magnetization_3D_T150} ($T=1.5,L=10$).
The radius of the blue sphere on Figure~\ref{Fig:Magnetization_3D_T150}
represent a residual value of $\langle M \rangle$ (see also Figure~\ref{ConvLatticesABCD}
bellow).
Note that the ellipsoidal character of the total 
spin vector distribution in the disordered phase
(Figure~\ref{Fig:Magnetization_3D_T150}) is merely an artifact of the parametrization (2).
 Calculations show that the distribution 
(Figure~\ref{Fig:Magnetization_3D_T150}) reduces to spherical form with 
temperature increase.

\begin{figure}[ht]
\centering
\includegraphics[width=6in]{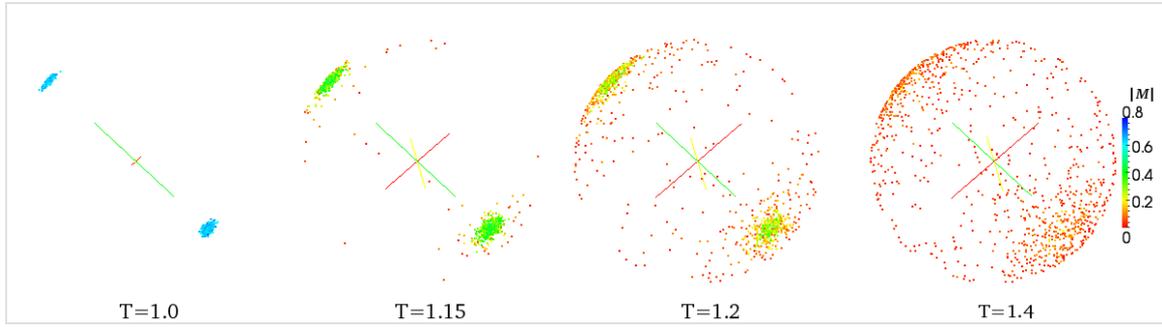}
\caption{(Color online) 
Distribution of normalized total spin  ($\bm{M}/|\bm{M}|$) for different temperatures.
Colors represent magnitude of the vector $\bm{M}$. (L=10)
}
\label{Fig:Magnetization_3D_sphere}
\end{figure}

To get a better insight into the angle distribution of total spin vectors,
we plot $\bm{M}/|\bm{M}|$ for various temperatures in Figure~\ref{Fig:Magnetization_3D_sphere}.
This figure also gives  complete information of total spin vectors $\bm{M}$ 
since the color  represents the magnitude of $\bm{M}$.

\begin{figure}[ht] 
\centering
\includegraphics [scale=0.25]{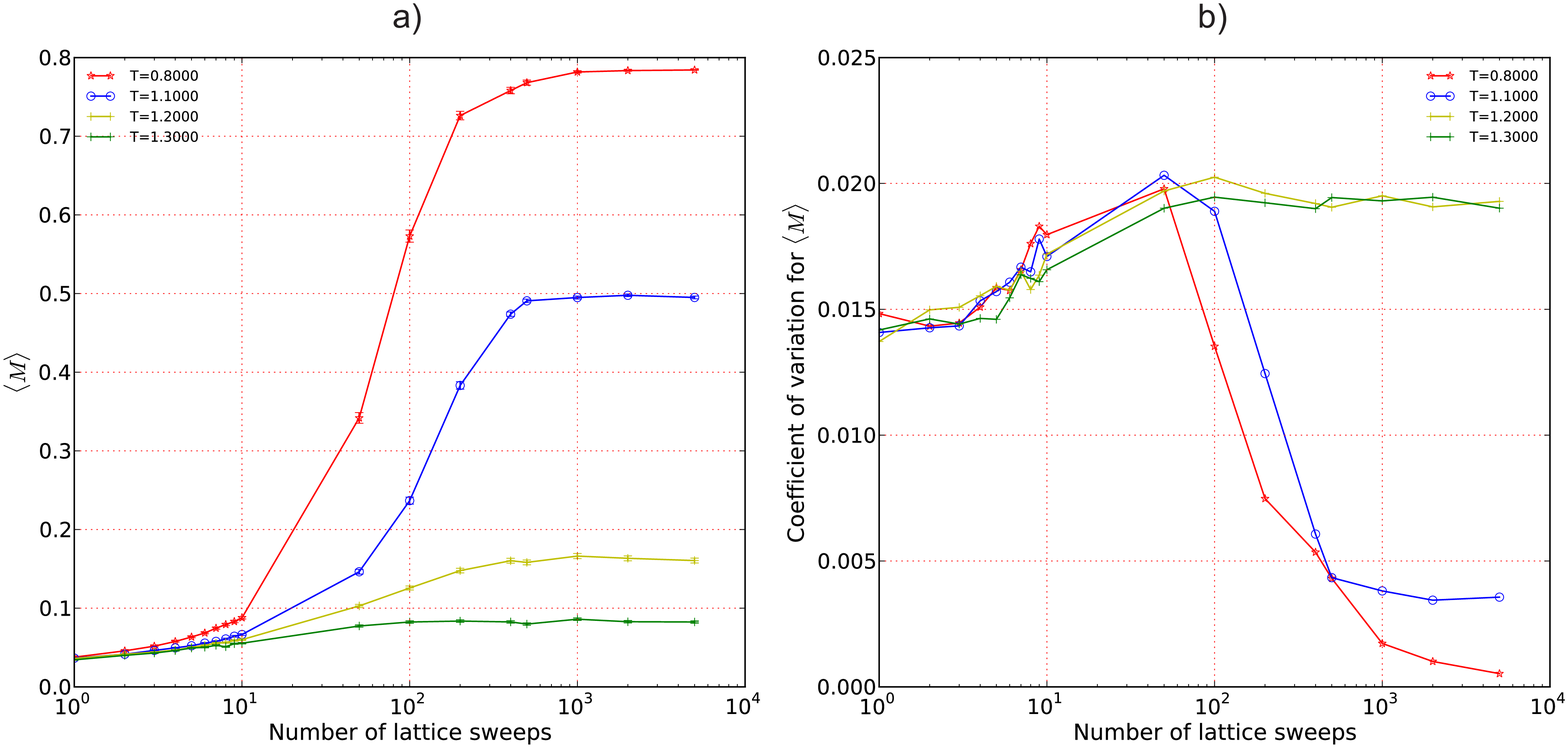}
\caption{(Color online) $L=10$; 
a) The order parameter $\langle{M}\rangle$ 
against number of LSs for $\mathcal{N} = 1000$ SPs for different temperatures ---
number of LSs required to reach equilibrium decreases with increase of temperature;
b) coefficient of variation (\ref{eq:cv}) of $\langle{M}\rangle$ against number of LSs.
When not shown, the error bars are smaller than  symbols.}
\label{Fig:TC_on_T}
\end{figure}

The order parameter $\langle{M}\rangle$ for the lattice with 10 lattice sites
in each dimension ($L=10$), on different temperatures is plotted against 
the number of LSs (Figure~\ref{Fig:TC_on_T}a).
Each simulation is conducted on $1000$ simulation paths.
The resulting figure exhibits two distinct parts.
In the first part magnetization rises rapidly with an increase in the number of LSs.
This part of the figure represents states that are walked trough 
while most of the lattices are still far from the thermal equilibrium state, 
also called 'the warm-up phase'.
The second part is based on representative 
or close--to--representative states, thus $\langle{M}\rangle$ 
in this part fluctuates mildly around the true value.
In our experiments the number of LSs is approximately doubled for each new simulation, 
so that details on the figure are much easier to observe if plotted on a 
semi--log plot, especially for the small LS values
(In contrast, a linear plot provides a better idea of thermalization duration 
and cost).
With the rise in temperature, $\langle{M}\rangle$ is diminishing, as expected.
The number of LSs necessary for almost all lattices to reach equilibrium is also reducing.
For $T=0.8$ thermalization takes $5 \times 10^3$ LSs but for $T=1.3$ only $500$ LSs is required.

Data set variability can be expressed in a number of different ways.
Usually variance or standard deviation are used. 
Both of these values are related to the data set mean and as such 
cannot be used for comparison of data sets with different mean values.
When such comparison is needed, coefficient of variation (cv) can be used.
Care should be taken because cv is meaningful only for data sets with positive means.
Coefficient of variation of a data set mean is given by:
\be
cv_{\bar{X}}=\frac{\sigma_{\bar{X}}}{\bar{X}}=
             \frac{\sigma_X}{\sqrt{N} \bar{X}}
\label{eq:cv}
\ee
\noindent
where 
$X$ represents any data set, 
$N$ number of elements in the set $X$,
$\bar{X}$ the data set mean,
$\sigma_{X}$ standard deviation of the set $X$
and
$\sigma_{\bar{X}}$ stands for standard deviation of the set mean $\bar{X}$.

Coefficients of variation of the $\langle{M}\rangle$ against number of LSs
are plotted on the Figure~\ref{Fig:TC_on_T}b), also for $1000$ simulation paths.
The figure shows that as simulation goes deeper in the disordered phase 
SO dispersion is bigger thus an increase in the coefficient of variation can be observed.
In order to get the same precision at higher temperatures 
it is necessary to compute more simulation paths.

\begin{figure}[H]
\centering
\includegraphics [scale=0.20]{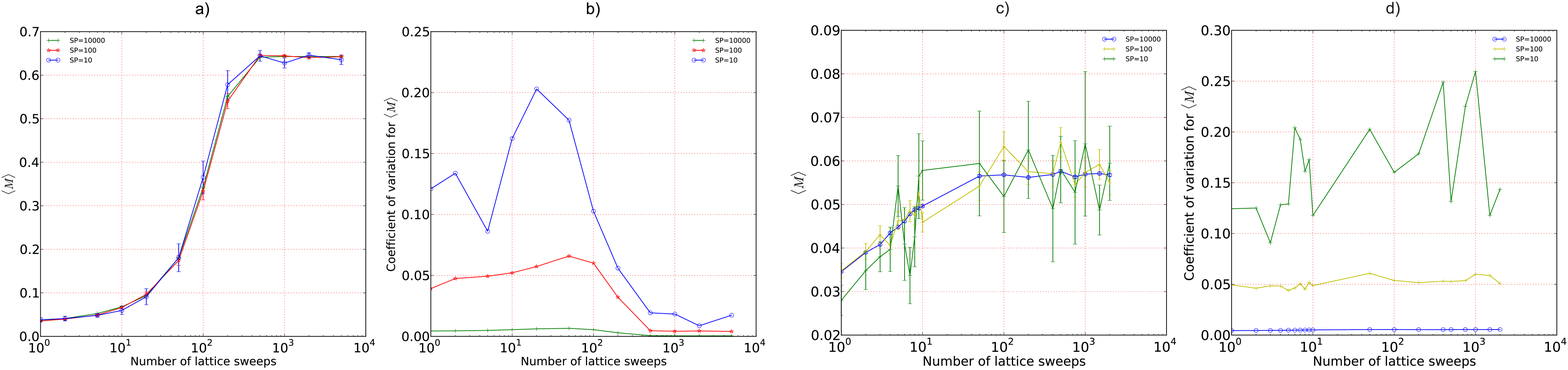}
\caption{(Color online) 
a) The order parameter $\langle{M}\rangle$ and b) the coefficient of variation 
(\ref{eq:cv}) 
(ordered phase, $L=10, T=1.0$) together with c) the order parameter
and d) the coefficient of variation (disordered phase, $L=10, T=1.5$)
 against number of LSs  for different number of SPs ($\mathcal{N}$). 
 When not shown, the error bars are smaller than  symbols. }
\label{L10T1_T150_M_cv}
\end{figure}

The order parameter $\langle{M}\rangle$ 
and the coefficient of variation of the order parameter mean
for the $L=10$ lattice in the ordered phase ($T=1.0$) for a different number of SPs 
are plotted against the number of LSs are shown on Figure~\ref{L10T1_T150_M_cv} a)
and Figure~\ref{L10T1_T150_M_cv}b).
When most of the path--chains reach equilibrium (around $500$ LSs) 
the coefficient of variation becomes quite small (just a couple of percents) 
even if only a couple of paths are used.
That is, most of the intensities  $\bm{M}_i$ entering (\ref{M_and_H})  
get very close to their mean value $\langle{M}\rangle$, thus
producing quite accurate and precise outputs.
If just an approximate result is needed, a simulation with just a few paths 
can be used to quickly and cheaply produce a reasonable estimate.
A similar plot, this time for the disordered phase, is shown in Figure~\ref{L10T1_T150_M_cv} c)
and Figure~\ref{L10T1_T150_M_cv} d).
In the disordered phase there is a much bigger dispersion due to significant 
thermal disturbances.
Thus, the coefficient of variation is larger compared to the ordered phase, 
even  after thermalization equilibrium has been reached 
(around $100$ LSs for $L=10$ and $T=1.5$).

$\langle{M}\rangle$ and the coefficient of variation of the $\langle{M}\rangle$ 
are shown against number of LSs conducted in ordered phase ($T=1.0$) for 
different lattice sizes (Figure~\ref{ConvLatticesABCD}).
Each simulation is performed on 1000 SPs.
Larger lattices generate more accurate (Figure~\ref{ConvLatticesABCD}a) and 
more precise (Figure~\ref{ConvLatticesABCD}b) output but require 
more LSs to reach a representative state (i.e.\ to reach thermal equilibrium).
Similar plots, just for the disordered phase ($T=1.5$), are also given in 
Figure~\ref{ConvLatticesABCD}. The inset on Figure~\ref{ConvLatticesABCD}a)
shows a close-up of the region in which the value of order parameter 
converges.
With an increase of lattice size accuracy improvements can be noticed, 
as $\langle{M}\rangle$ is approaching zero  
(Figure~\ref{ConvLatticesABCD}c).
Since thermal disturbance is significant in the disordered phase
there is no increase in precision (Figure~\ref{ConvLatticesABCD}d).

\begin{figure}[hbt] 
\includegraphics [width=13.0cm]{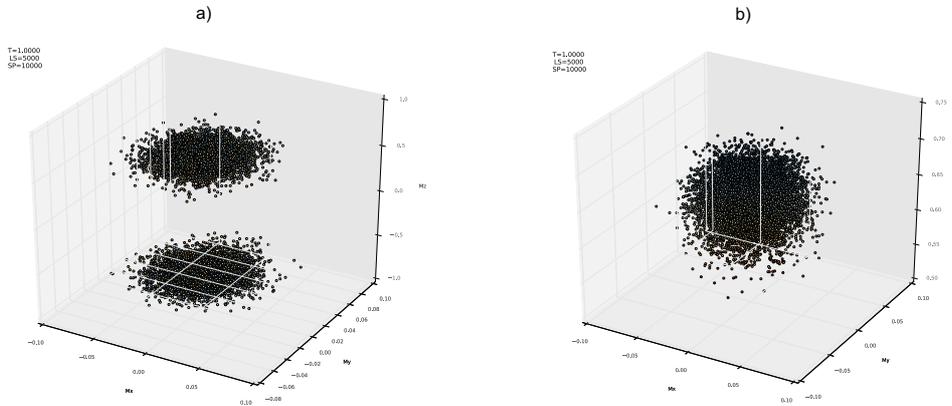}
\centering
\caption{(Color online) Distribution of total spin 
at $T=1$, for $5\times 10^3$ lattice sweeps and $\mathcal{N} = 10^4$ simulation paths.  Every path
started a) from random spin configuration and b) from ordered configuration.}
\label{Fig:goredole}
\end{figure}

\begin{figure}[ht] 
\includegraphics [scale=0.30]{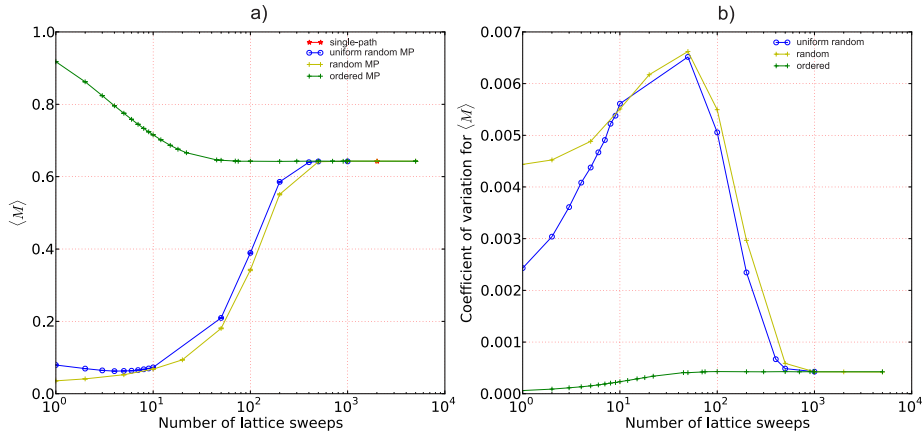}
\centering
\caption{(Color online) a) Magnetization and b) coefficient of variation for $L=10$, $T=1$, calculated starting
from both ordered and disordered states, as a function
of number of lattice sweeps for $\mathcal{N}=10000$. Red dot represents single-path result.
When not shown, the error bars are smaller than the symbols.}
\label{Fig:ee}
\end{figure}

The number of lattice sweeps needed for a lattice to reach it's representative state 
(also called burn-in or warm--up phase) is unknown.
It depends on many parameters and can vary substantially.
Insufficient number of lattice sweeps causes inaccurate simulation results.
To overcome this problem for each temperature
half of the simulation paths are computed from the random initial state 
where other half started from the ordered state
(see Figure \ref{Fig:goredole}) 
These two sets are averaged using (\ref{MCAV}) but results from each half separately.
When both halves produce the same result (see Figure \ref{Fig:ee}) we can be reasonably certain
that it is an accurate value.

At this point we make  connection with standard (single-path) simulation
and show that the states generated by  each path  are truly representative ones. To do so, we
examine the value of the ordered parameter for $L=10$ and $T=1$ 
(see Figure \ref{Fig:ee}) and show that single-path and
MP results converge to the same value. The single path result 
 obtained from $10^4$ measurements is represented by red
star on Figure \ref{Fig:ee} and the  error,
estimated by the blocking method, is less than the size of symbol.
The ordered state is used as a starting point in the single-path simulation.
On the other hand, MP value is obtained from $10^4$ simulation paths.
Three sets of data correspond to three different initial configurations:
ordered state (green symbols), the initial state with all angles
taking arbitrary values (yellow symbols) and the uniform state where
unit vectors are uniformly distributed on the sphere (blue symbols).
It is seen that, within the error which is $\sim 10^{-4}$, all of them
converge to the same value of the order parameter. Thus, we conclude
that the final states on each path are indeed representative ones.
Moreover, the Figure~\ref{L10T1_T150_M_cv}a) shows that the value
of $\langle M \rangle$, obtained with just several hundreds of
paths comes very close to the convergence point.
All of this should not come as a surprise, since each path  represents 
one Markov chain whose trajectory through the phase space is governed by
the Metropolis algorithm.

\section{Thermodynamics of classical Heisenberg model}\label{SectResult}

\begin{figure}[ht] 
\includegraphics [width=15.0cm]{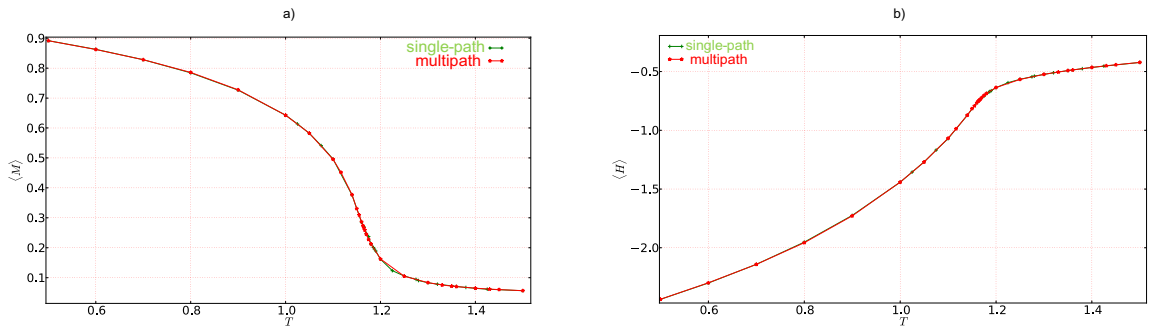}
\centering
\caption{(Color online) a) Magnetization and b) energy as  functions of temperature for $L=10$ in the single--path
and multipath approach.}
\label{Fig:mag}
\end{figure}

\begin{figure}[ht] 
\includegraphics [width=15.0cm]{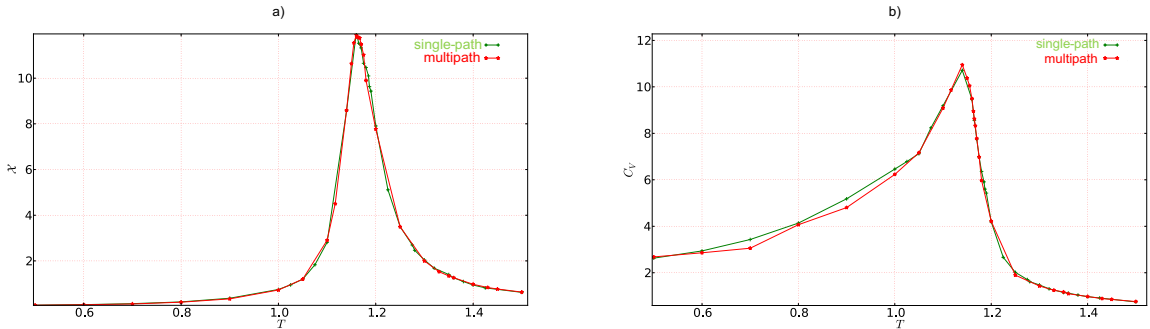}
\centering
\caption{(Color online) a)Magnetic susceptibility and b) heat capacity as  functions of temperature for 
$L=10$ in the single--path and multipath approach.}
\label{Fig:susc}
\end{figure}

Further justification of the MP approach comes from the study of thermodynamic
properties of CHM over wide temperature range. We examine spontaneous
magnetization (the ordered parameter), energy, susceptibility and
specific heat per lattice site and compare results from MP and single-path
simulation. All simulations were conducted for the system with $N=10^3$ lattice sites  and
periodic boundary condition, both in single and multipath approach.
In single--path approach we used $2\times 10^6$ lattice sweeps to achieve
thermal equilibrium in whole temperature range, and afterwards  one out
of every five lattice sweeps was used to calculate the averages of
physical quantities \cite{Kitaev}. At every temperature $5 \times 10^5$ measurements were averaged.
To make sure that reliable results are generated by multipath simulation, it is prepared
in two different setups. In the first one, refered to as random initial state simulation in the text,
at every lattice site both angles $\theta$ and $\varphi$ are taken to be arbitrary. In the second one,
denoted as ordered initial state simulation spins are taken to points along z-axis, with no restriction
on second spherical angle $\varphi$. The MP simulation is conducted with 
$\mathcal{N}= 10^4$ SPs.
The results are shown on Figures \ref{Fig:mag} and \ref{Fig:susc}
and they reveal that the differences 
in the thermodynamical characteristic
obtained by single--path and multipath 
approach are negligible.
The analysis from previous section indicates that the increase
in the number of lattice sweeps, simulation paths and the 
lattice sites, leading to the enhancement in precision and 
accuracy,   should  reduce this difference even further.
We have to bear in mind, however, that multipath simulations naturally split
 into three temperature domains in which different numbers of lattice sweeps/simulation paths are needed.
In low temperature region for simulation convergence
more lattice sweeps is needed since all paths start
from some random state of the lattice.
(Simulation speed can be optimized if 
ordered state is taken to be ''starting point'' of all paths.) 
On the other hand, high temperature region
 requires more simulation paths. In the critical region
we take sufficiently large number of lattice sweeps and results due
to overlapping of the distinct output distributions \cite{Binder81}.

\begin{figure}[ht] 
\includegraphics [scale=0.5]{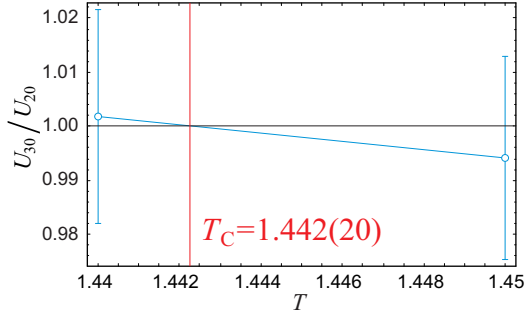}
\centering
\caption{(Color online) Determination of the critical temperature: the ratio of 
Binder cumulants $U_{30}$ and $U_{20}$ calculated for $T=1.44$ and $T=1.45$ (blue symbols)}
\label{Fig:TCfig}
\end{figure}

As another illustration of  the multipath approach, we calculate
the critical temperature of $O(3)$ CHM. 
It is estimated from the condition $U_{L}(T_{\tm C}) /U_{L'}(T_{\tm C}) = 1$
for $L = 30$ and $L'=20$. Simulations for each lattice size on $T=1.44$ and $T=1.45$ were conducted using 
$10^{6}$ SPs and  200 lattice sweeps. The results are shown
on Figure~\ref{Fig:TCfig} and the final result is 
$T_{\tm C} = 1.442 (20)$ i.e. $\beta_{\tm C} = 0.693 (9)$. The standard 
estimates for the inverse critical temperature of CHM, are 
$\beta_{\tm C} = 0.69300(10)$ \cite{Janke},  $\beta_{\tm C} = 0.693035(37)$ \cite{Kun} and
$\beta_{\tm C} = 0.693002(12)$ \cite{Ballesteros} (se also \cite{Butera}). Thus, the current MP
simulation, with only 200 lattice sweeps, reaches three-digit precision for $\beta_{\tm C}$.
One should note that the values quoted from papers \cite{Janke, Kun, Ballesteros}
are obtained  using cluster algorithms, larger lattices  (up to $L=48$)
and with significantly more laattice sweeps (For example, the authors of \cite{Kun}
used $10^6$ measurments after  $1-7 \times 10^4$ single-cluster updatings).
Nevertheless, the results from Section \ref{SectSimOut} indicate that 
the more precise value for $T_{\tm C}$ (i.e.  $\beta_{\tm C}$) is expected
when MP simulation is conducted on larger lattices  with more lattice sweeps
and simulation paths. Of course, the shortening of each single chain 
(i.e.\ shortening of sequentially dependent algorithm parts)
compared to an increase in total number of LSs (i.e.\ increase in total number of processor cycles utilized)
is obviously a trade-off and depends on the problem in question.
The detailed study of critical behaviour of CHM, employing MP technique, is a work in progress.

\section {Conclusions}

The Metropolis algorithm applied to multiple random--walk paths
becomes an embarrassingly parallel algorithm in which
plenty of cores can be utilized easily. Multipath approach produces normally distributed 
simulation output with an easily computable error margin.
Also, this approach is local minima entrapment resilient by definition.
Multipath approach is based on a slightly modified Metropolis algorithm as 
it generates just one representayive state
after a number of thermalization sweeps, conducted for warm--up. 
Therefore Markov chain on each simulation path is significantly shorter 
than the chain in the equivalent single--path simulation.
Usually path lengths differ by a couple of orders of magnitude, 
offering tremendous parallelization speedups.

The present paper is mainly devoted to  investigation of the order 
parameter (spontaneous magnetization). While results from first half of Section \ref{SectSimOut}
demonstrate the quality of output data, remaining ones from \ref{SectSimOut} and
those from Section \ref{SectResult}  reveal, beside inherent normal distribution
of output results,  two more important aspects
of MP simulation of CHM. First, each path generates representative states,
since MC average  with just several hundreds
of simulation paths (or less) yields reasonable estimate of the order parameter.
Second, MC averages of the order parameter calculated in MP and single-path
simulation coincide (within errorr bars) over wide temperature range.
On the other hand, the quantitative description of the critical region (high-precision
determination of the critical temperature and critical indices), recquires 
a similar analysis for higher-order moments.

\section*{Acknowledgments}
This work was supported by the Serbian Ministry of
Education and Science under Contract No. OI-171009.
The authors acknowledge the use of the Computer Cluster of the
Galicia Supercomputing Centre (CESGA).

\appendix

\section{2D Ising model}\label{ising}
\begin{figure}[htb] 
\includegraphics[scale=0.35]{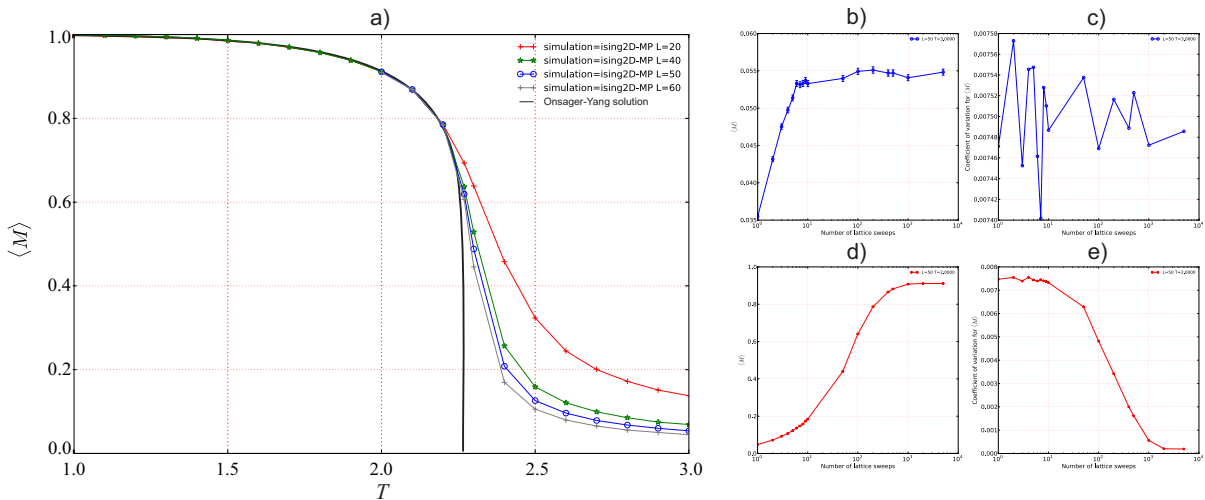}
\centering
\caption{(Color online) 
Results of MP simulations for Ising model on square lattice. The parameters of simulations
are given in the text.}
\label{Fig:Izing}
\end{figure}

To test the applicability of MP simulation on anisotropic systems, 
we examine it on 2D Ising model. The exact solution of this model is known, 
so MP simulations can be checked against Onsager--Jang result.
As in the case of CHM, we shall focus on the order parameter.

The setup of the MP simulation is the same as in the case of $O(3)$ CHM. 
This means that each path start from random state of the lattice and
contributes to MC averages with its final state.  The Figure~\ref{Fig:Izing}
displays MP results for spontaneous magnetization on latices
$20 \times 20, 40 \times 40, 50 \times 50$ and $60 \times 60$.
Each simulation is conducted with $3\times 10^4$ lattice sweeps
and $10^4$ SPs. Specificaly, Figure~\ref{Fig:Izing}a) shows the order
parameter as a function of temperature for several lattices, together
with Onzager-Yang solution \cite{OnzagerYang}. As expected, simulations
on larger lattices produce more accurate values of spontaneous
magnetization. Further, the convergence
of MC average of the ordeder parameter and corresponding coefitient of variation
against number of lattice sweeps are shown  in para--phase [$T=3$, Figures 
~\ref{Fig:Izing}b) and c)] and in ferro--phase [$T=1$, Figure~\ref{Fig:Izing}d) and e)].
All graphs from Figure ~\ref{Fig:Izing} ilustrate that order parameters
of models with continous (CHM) and discrete (IM) symmetry behave
similarly in MP simulation. Thus, the entire analysis and all conclusions from Sections
\ref{SectSimOut} and \ref{SectResult} apply almost without
any change to IM. Finally, we determine the critical temperature of  IM
on square lattice by method of Binder cumulants. By using lattices 
$50 \times 50$ and $60 \times 60$ for $T=2.264$ and $T=2.268$ in MP
simulation with $3 \times 10^4$ lattice sweeps  and $5\times10^4$ SPs,
we obtain $T_{\tm{C}} = 2.269(10)$. This value agrees with exact solution
in three decimal places
As in the case of CHM, a simulation on larger lattice and with more lattice sweeps and 
SPs is nedeed for more precise value of the critical temperature.

\bibliographystyle{elsarticle-num}
\bibliography{paper}

\end{document}